\documentclass[journal]{new-aiaa}
\usepackage[utf8]{inputenc}
\usepackage{textcomp}

\usepackage{graphicx}
\usepackage{amsmath}
\usepackage[version=4]{mhchem}
\usepackage{siunitx}
\usepackage{longtable,tabularx}
\usepackage{blindtext}
\RequirePackage[capitalise]{cleveref}
\usepackage{mymacros}
\usepackage[ruled,vlined]{algorithm2e}
\RestyleAlgo{ruled}
\SetKwComment{Comment}{/* }{*/}
\newcommand\ColoredComment[1]{{\color{sntGray} \Comment{#1}}}
\renewcommand{\arraystretch}{0.9}
\usepackage{mathtools}
\usepackage{nicefrac}
\usepackage{color,soul}
\newcounter{problem}
\newtheorem{Problem}[problem]{Problem}
\usepackage{multirow}
\usepackage{xcolor, colortbl}

\newcommand{\arraystretchfortable}{1.2}
\newcommand{\arraystretchformatrixintable}{1}
\newcommand{\hseparationofmatrixintable}{2pt}

\usepackage{setspace}

\title{Fuel-Optimal Formation Reconfiguration by Means of Unidirectional Low-Thrust Propulsion System}

\def\sntAffil{SnT, University of Luxembourg, 29, Avenue J.F Kennedy, 1855, Luxembourg, Luxembourg}
\def\polimiAffil{Department of Aerospace Science and Technology, Politecnico di Milano. 34, via La Masa, 20156, Milan, MI, Italy}
\def\LuxSpaceAffil{LuxSpace, 9, Rue Pierre Werner, 6832 Betzdorf, Luxembourg}
\def\FSTMAffil{Faculty of Science, Technology and Medicine, University of Luxembourg, 2, place de l’Université, 4365, Esch-sur-Alzette, Luxembourg}

\def\firstAuthorFN{PhD researcher, Ahmed.mahfouz@uni.lu}
\def\secondAuthorFN{Assistant Professor,  Gabriella.gaias@polimi.it}
\def\thirdAuthorFN{Senior Engineer, dallavedova@luxspace.lu}
\def\fourthAuthorFN{Full Professor, Holger.Voos@uni.lu}

\author{Ahmed Mahfouz \footnote{\firstAuthorFN}}
\affil{\sntAffil}
\author{Gabriella Gaias \footnote{\secondAuthorFN}}
\affil{\polimiAffil}
\author{Florio {Dalla Vedova} \footnote{\thirdAuthorFN}}
\affil{\LuxSpaceAffil}
\author{Holger Voos \footnote{\fourthAuthorFN}}
\affil{\sntAffil, \\ \FSTMAffil}

\begin{document}
\maketitle

\section{Introduction}
\lettrine{L}{ow-thrust} electric propulsion systems are becoming the preferred design solution for small-class and CubeSat-based satellite platforms \cite{Miller2021Survey}. This is motivated
by many factors, among which is the fact that they are generally more fuel efficient than their chemical counterparts. Furthermore, electric thrusters are  require significantly less propellant mass which leads to lighter launch mass and consequently reduces the costs \cite{Krejci2018SpacePropulsion}. To further minimize the complexity of the system design and/or to meet stringent power/mass constraints of small satellite platforms, the propulsion system typically features a single throttleable electric thruster. Examples of such satellites include Triton-X Medium and Heavy \cite{Helmeid2022TheIntegrated}, the PLATiNO platform \cite{Stanzione2018PLATiNO}, and the Gravity Field and Steady-State Ocean Circulation Explorer (GOCE) satellite \cite{floberghagen2011mission}. Small-size satellites can serve several multi-satellite missions, such as cooperative formation-flying to build disaggregated instruments and rendezvous activities to carry out inspection missions. These demand for the capability to perform formation reconfiguration by means of low-thrust unidirectional propulsion systems. In this note, the term "unidirectional" signifies that a the propulsion system comprises a single un-gimbaled thruster.\\

So far the relative orbit reconfiguration problem has been widely addressed. Examples include guidance and control schemes for formations that employ impulsive thrusters 
\cite{Larsson2011Autonomous, Gaias2015Impulsive, di2018continuous, chernick2018new}, however, these approaches are not viable for formations that utilize electric thrusters. Other guidance approaches were proposed for the low-thrust case, e.g. for the Formation Flying L-band Aperture Synthesis (FFLAS) mission study in \cite{scala2021design}, and for collision avoidance maneuver optimization \cite{DeVittori2022Low-Thrust}. A core assumption in these guidance schemes is the omnidirectional thrusting capability, which hinders the adaption of these approaches for the unidirectional thrusting case.
A unidirectional propulsion system was adopted in the Autonomous Vision Approach Navigation and Target Identification (AVANTI) mission \cite{Gaias2018Avanti} where the developed guidance scheme, which is only suited for impulsive thrust satellites, allowed the re-orientation of the thruster nozzle through planned control windows \cite{Gaias2015Impulsive, Gaias2015Generalized}.
Furthermore, Model Predictive Control (MPC) schemes were proposed in \cite{Belloni2023Relative, Mahfouz2023Autonomous} to address the problem of formation reconfiguration and formation keeping for satellites which are equipped with single electric thrusters, yet the mission operational constraints, i.e. necessary thruster off-periods, were not considered.\\

In this note, the problem of autonomous optimal formation reconfiguration is addressed for a formation which comprises two satellites, a chief and a deputy, where only the deputy has orbit maneuvering capabilities through a single throttleable electric thruster. In this setting, an attitude slew maneuver is necessary before each thruster firing so that the nozzle could be aligned with the required thrust direction before the firing takes place. The problem is approached through formulating a trajectory optimization (guidance) problem as a fuel-optimal constrained convex optimization problem with multiple no-thrust windows during which attitude redirection slews takes place. This, in-turn, requires constant alternation between the on and off states of the thruster.
One big advantage of this on-off alternation is the ability to accommodate long no-thrust periods which arise form mission constraints, e.g. eclipse, during which electric thrusters are usually turned off since the solar arrays are not generating electricity. 
The biggest advantage of the proposed guidance scheme is that, by relaxing some of the original constraints, it can be transformed into a Quadratic Programming (QP) problem which can be solved efficiently using any of the standard QP solvers. This makes it a attractive candidate to be implemented onboard of a satellite that uses one of the Commercial-Off-The-Shelf (COTS) onboard computers.
The control loop is then closed through an MPC-like scheme where the optimization of the state and input thrust profiles (from the current to the final time) are optimized using the proposed guidance algorithm. It is to be noted that the control logic does not necessitate the guidance algorithm to be run at each MPC step, but rather uses the previous guidance profile if the current state is close, within a predefined threshold, to its value predicted by the previous guidance solution.
The main contributions of this note are a) An efficient-to-solve guidance scheme for formation reconfiguration when the controlled satellite is equipped with a single electric thruster; b) The ability of the guidance scheme to accommodate as many long no-thrust periods as the user dictates; c) An MPC-like algorithm to close the control loop which does not require the guidance to be run at each receding horizon.
Note that in the remainder of the text, the terms "fuel-optimal" and "$\Delta V$-optimal" are used interchangeably. These two terms are identical only in the case of single-directional propulsion systems \cite{ROSS2006SpaceTrajectoryOptimization}.\\


This Note presents applications where collision avoidance is not of a concern for the satellite reconfigurations. Nonetheless, if required by a specific application, the proposed methodology is straightforward applicable by adding the collision avoidance constraints through the relaxed formulation generally adopted in the literature \cite{Morgan14MPC}. By including such affine constraints, in fact, the overall optimization problem is still cast into the QP formulation, thus featuring all the proposed advantages.\\


This research comes as part of the AuFoSat  toolbox to support the future missions Triton-X; a multi-mission microsatellite platform developed by LuxSpace to accommodate various types of payloads in Low Earth Orbits (LEO). Previous AuFoSat research discussed orbit design \cite{menzio2022formation}, relative navigation \cite{mahfouz2022relative, Mahfouz2023GNSS-based}, and absolute orbit keeping for Triton-X \cite{Mahfouz2023Autonomous}. While the developed algorithms in the framework of AuFoSat are meant to primarily be used onboard of Triton-X, the goal is to provide a guidance, navigation, and control toolbox that could be used by any satellite with the same, or similar, specifications as Triton-X.\\

This article is organized such that the following section of introduces the formulation of the relative orbital dynamics using the Relative Orbital Elements (ROE) formulation. In \cref{sec:Guidance}, the guidance problem is formulated as a convex programming problem, then it is transformed to a QP problem for it to be solved efficiently. The guidance scheme is validated in \cref{sec:Guidance_validation} where it is shown to work efficiently in the presence as well as in the absence of long no-thrust periods. Finally the module execution logic is introduced in \cref{sec:Closed_loop} and the closed loop system is benchmarcked against another controller from the literature which deals with a similar problem.\\

\section{Dynamical model}

The reference frames used in this work are; the Earth-Centered-Inertial frame (ECI), denoted as $\set{F}^{i}$, the Satellite-body-fixed frame, denoted as $\set{F}^{b}$, and the Radial-Transversal-Normal frame (RTN), denoted as $\set{F}^{r}$. The reader is advised to refer to \cite{Mahfouz2023Autonomous} for a full definition of these reference frames. Vectors expressed in $\set{F}^{i}$, $\set{F}^{b}$, or $\set{F}^{r}$ are signified by the superscripts $\parenth{\cdot}^{i}$, $\parenth{\cdot}^{b}$, or $\parenth{\cdot}^{r}$ respectively.\\

The orbital motion of a satellite under the gravitational influence of a major body (e.g. the Earth) can be parameterized in a planet-centered inertial frame by the following set of orbital elements,
\begin{equation} \label{eq:OE}
    \vec{\alpha} \coloneqq \begin{bmatrix} a & u & e_{x} & e_{y} & i & \Omega \end{bmatrix}^{\intercal}, 
\end{equation}
where $a$ is the semi-major axis, $u$ is the mean argument of latitude, $\vec{e}\coloneqq\begin{bmatrix} e_{x} & e_{y} \end{bmatrix}^{\intercal} = \begin{bmatrix} e \cos{\omega} & e\sin{\omega} \end{bmatrix}^{\intercal}$, is the eccentricity vector with $e$ being the orbital eccentricity, $i$ is the orbital inclination, and $\Omega$ is the Right Ascension of the Ascending Node (RAAN). It is important to note that the motion of the satellite can also be parameterized by the Cartesian state vector, $\vec{x}^{i}\coloneqq \begin{bmatrix} \parenth{\vec{r}^{i}}^{\intercal} & \parenth{\vec{v}^{i}}^{\intercal}
\end{bmatrix}^{\intercal}$, where $\vec{r}^{i}$ and $\vec{v}^{i}$ are the absolute position and velocity vectors expressed in $\set{F}^{i}$ which can be mapped to/from the orbital elements through a set of nonlinear equations \cite{vallado2001fundamentals}. The exact position and velocity of the spacecraft transform into osculating orbital elements, which in the remainder of this work will be denoted by $\tilde{\vec{\alpha}}$. Mean orbital elements, denoted by $\vec{\alpha}$, are to be intended as one-orbit averaged elements, where the short- and long-term oscillations generated by the $J_{2}$ harmonic of the Earth gravitational potential are removed. Mean/osculating elements' conversions are performed through the transformations developed in \cite{Gaias2020Analytical}.\\
  
The relative motion between a deputy and a chief spacecraft can be described by the dimensionless quasi-nonsingular Relative Orbital Elements (ROE) vector which is a nonlinear transformation of the orbital elements vector introduced in \cref{eq:OE},
\begin{equation} \label{eq:ROE}
    \delta \vec{\alpha} \coloneqq \begin{bmatrix} \delta a & \delta \lambda & \delta e_{x} & \delta e_{y} & \delta i_{x} & \delta i_{y} \end{bmatrix}^{\intercal} = \begin{bmatrix} \Delta a/a_{c} & \Delta u + \Delta \Omega \cos{i_{c}} & \Delta e_{x} & \Delta e_{y} & \Delta i & \Delta \Omega \sin{i_{c}} \end{bmatrix}^{\intercal},
\end{equation}
where $\delta \vec{\alpha}$ is the dimensionless ROE vector, $\delta a$ is the relative semi-major axis, $\delta \lambda$ is the relative mean longitude, $\delta\vec{e}\coloneqq\begin{bmatrix} \delta e_{x} & \delta e_{y}\end{bmatrix}^{\intercal}$ is the relative eccentricity vector, and $\delta\vec{i}\coloneqq\begin{bmatrix} \delta i_{x} & \delta i_{y}\end{bmatrix}^{\intercal}$ is the relative inclination vector. It is to be noted that here, and in the coming discussions, the subscript $\parenth{\cdot}_{d}$ denotes a quantity related to the deputy satellite, while the subscript $\parenth{\cdot}_{c}$ is used for chief-related quantities. Moreover, $\delta \parenth{\cdot}$ signifies a relative quantity between the deputy and the chief which is not necessarily the arithmetic difference between that of the deputy and that of the chief, while $\Delta \parenth{\cdot}$ signifies the arithmetic difference between $\parenth{\cdot}_{d}$ and $\parenth{\cdot}_{c}$, i.e.  $\Delta \parenth{\cdot} \coloneqq \parenth{\cdot}_{d} - \parenth{\cdot}_{c}$. As in the case of absolute orbital elements, the osculating ROE vector is denoted by $\delta \tilde{\vec{\alpha}}$, whereas the mean ROE vector is referred to as $\delta {\vec{\alpha}}$. A dimensional ROE vector is obtained by multiplying the dimensionless ROE vector by the semi-major axis of the chief,
\begin{equation} \label{eq:y_definition}
    \vec{y} = a_{c} \delta \vec{\alpha},
\end{equation}
where $\vec{y}$ is the dimensional mean ROE vector with units of length.


Assuming neighbouring orbits of the chief and the deputy, and a near-circular orbit of the chief, the dynamics of the ROE can be linearized to the first order considering the mean effect of the $J_{2}$ zonal harmonic. In fact, a closed form solution of the linearized dynamics can be obtained for piece-wise constant input acceleration as discussed in \cite{di2018continuous}. The system evolution is expressed in the following form,
\begin{equation}\label{eq:ROE_dynamics_sol}
    \vec{y} \parenth{t_{k+1}} = \mat{\Phi}\parenth{t_{k}, t_{k+1}} \vec{y} \parenth{t_{k}} + \frac{a_{c}}{M} \mat{\Psi}\parenth{t_{k}, t_{k+1}} \vec{f}^{r}\parenth{t_{k}, t_{k+1}}, 
\end{equation}
where $\mat{\Phi}\parenth{t_{k}, t_{k+1}}$ is the State Transition Matrix (STM) between the two time instants, $t_{k}$ and $t_{k+1}$, $\mat{\Psi}\parenth{t_{k}, t_{k+1}}$ is the convolution matrix between the same two time instants, $M$ is the deputy's mass which is assumed constant throughout any maneuver, and $\vec{f}^{r}\parenth{t_{k}, t_{k+1}} = \begin{bmatrix} f_{R} & f_{T} & f_{N} \end{bmatrix}^{\intercal}$ is the input thrust vector in $\set{F}^{r}$; constant over the period $\left[t_{k}, t_{k+1}\right)$. In the rest of the text, and in order to simplify the representation of equations, the following notations are used, $\mat{\Phi}_{k\vert k+1} \equiv \mat{\Phi}\parenth{t_{k}, t_{k+1}}$, $ \mat{\Psi}_{k\vert k+1} \equiv \mat{\Psi}\parenth{t_{k}, t_{k+1}}$, $\vec{y}_{k} \equiv \vec{y}\parenth{t_{k}}$, and $\vec{f}^{r}_{k\vert k+1} \equiv \vec{f}^{r}\parenth{t_{k}, t_{k+1}}$.

\section{Guidance}\label{sec:Guidance}
\begin{figure*}[ht]
    \centering
    \includegraphics[width=\linewidth]{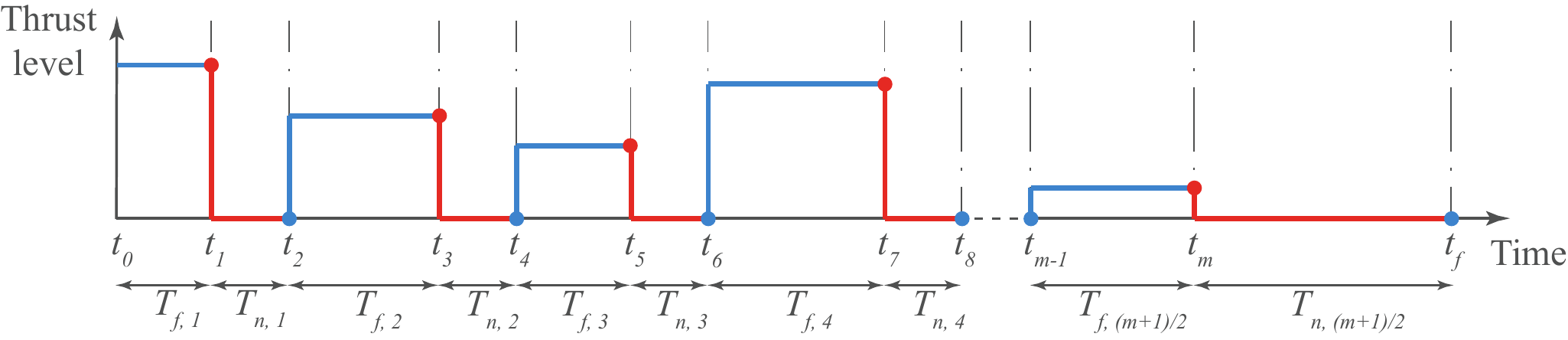}
    \caption{Graphical representation of the low-thrust guidance scheme}
    \label{fig:Low-thrust-guidance-scheme}
\end{figure*}

In this section, a multiple shooting guidance scheme is developed when large relative orbit maneuvers are required between two satellites, a deputy and a chief, where the chief might be either a physical satellite or a virtual one. 
The deputy satellite is assumed to be equipped with a single throttleable electric thruster, which not only mandates redirection slew maneuvers before every thruster firing, but also dictates the thruster to operate perpetually since it provides low thrust. It is for these reasons that the guidance scheme is designed from the beginning to operate on an alternating on/off mode where the throttleable thruster is turned off to allow the attitude maneuver to taking place.
The trajectory optimization problem is formulated such that the a change in relative orbit is required from $\vec{y}_{0}$ at $t_{0}$ to a reference $y_{f}$ at $t_{f}$ through
$\dfrac{m+1}{2}$ continuous thruster firings, where $m$ is an odd number. Figure \ref{fig:Low-thrust-guidance-scheme} illustrates the alternation between thrust and attitude maneuvers throughout the allocated maneuver time, from $t_{0}$ to $t_{f}$.
In order to enhance the predictability of the mission, the vector of time instants at which the thruster is turned on and off, $\vec{t} = \begin{bmatrix} t_{0} & t_{1} & \hdots & t_{f}
\end{bmatrix}$, are not treated as optimization variables and are left as a user input. In \cref{fig:Low-thrust-guidance-scheme}, $\forall l \in \dist{L} = \left\{1,\; 2,\; \hdots,\; \left(m+1\right)/2\right\}$, $T_{f, l}$ are the forced (thrust-powered) time periods, while $T_{n, l}$ are those during which the natural unforced translational dynamics take over (coast arcs).

\subsection{problem formulation}
Letting,
\begin{equation}\label{eq:Y_definition}
    \mat{Y} = \begin{bmatrix}
        \vec{y}_{0} &
        \vec{y}_{1} &
        \hdots &
        \vec{y}_{m+1}
    \end{bmatrix}, \quad \mat{F} = \frac{a_{c}}{M} \begin{bmatrix}
        \vec{f}^{r}_{0\vert 1} &
        \vec{f}^{r}_{1\vert 2} &
        \hdots &
        \vec{f}^{r}_{m\vert m+1} \end{bmatrix},
\end{equation}
the guidance problem can be formally written as an optimization problem as follows,
\begin{Problem}\label{prob:original_formulation}
\begin{align}
& \min_{\mat{Y}, \mat{F}} \quad \textnormal{tr}\parenth{\mat{F}^{\intercal}\mat{F}} \qquad \text{subject to,} \nonumber\\
& \vec{y}_{0} = \vec{y}_{0}, \qquad \vec{y}_{m+1} = \vec{y}_{f}, \qquad \vec{y}_{k+1} =  \mat{\Phi}_{k\vert k+1} \vec{y}_{k} + \frac{a_{c}}{M} \mat{\Psi}_{k\vert k+1} \vec{f}^{r}_{k\vert k+1} \quad \forall k \in \dist{K},\\
& \vec{f}^{r}_{k\vert k+1} = \vec{0} \quad \forall k \in \dist{K}_{n}, \label{eq:u0_constraint_original}\\
& \norm{\vec{f}^{r}_{k\vert k+1}} \leq f_\text{max} \quad \forall k \in \dist{K}_{f}, \label{eq:umax_constraint_original}
\end{align}
\end{Problem}
where $\dist{K} = \dist{K}_{f} \cup \dist{K}_{n}$, with $\dist{K}_{f} = \left\{0, 2, 4, \hdots, m-1\right\}$ and $\dist{K}_{n} = \left\{1, 3, 5 \hdots, m\right\}$, $\textnormal{tr}\parenth{\cdot}$ is the matrix trace, and $f_\text{max}$ is the maximum allowable thrust by the onboard thruster.
Note that \cref{eq:u0_constraint_original} is a hard constraint to assure there  is no input thrust provided during attitude redirection maneuvers. 
Indeed, Problem \ref{prob:original_formulation} stands as a Convex Optimization Problem (COP) since it fulfills all the necessary conditions for a problem to be one \cite{boyd2004convex}, which namely are; a) Cost function must be convex (for minimization problems); b) Inequality constraints must be convex; c) Equality constraints must be affine.
Being a COP, Problem \ref{prob:original_formulation} is guaranteed to have a unique solution, however, it would be much more efficient to solve if it could be put in one of the standard classes of convex optimization problems, e.g. Linear Programming (LP), Quadratic Programming (QP), etc. since dedicated solvers for these classes have matured over the past decades.
It is clear that the only thing which prevents Problem \ref{prob:original_formulation} from being put in the convex QP canonical form is constraint \eqref{eq:umax_constraint_original} since it is a quadratic constraint and not an affine one. It can be, nonetheless, transformed into multiple affine constraints using the methodology proposed in \cite{camino2019linearization} which suggests that,
\begin{equation}\label{eq:norm_constraint_transformation}
    \norm{\vec{b}} \leq d, \quad \vec{b} \in \set{R}^{2}\qquad
    \text{can be relaxed by} \qquad
    \begin{bmatrix} \cos{\parenth{\gamma_{j}}} & \sin{\parenth{\gamma_{j}}} \end{bmatrix} \vec{b} \leq d \cos{\parenth{\gamma_\text{max}}} \quad \forall j \in \dist{J},
\end{equation}
where $\dist{J} = \left\{1, 2, \hdots, n_\text{dir} \right\}$, with $n_\text{dir}\geq 3$ being the number of affine inequality constraints that approximate the Euclidean norm constraint, $\gamma_{j} = \dfrac{2\parenth{j-1}\pi}{n_\text{dir}} + \gamma_\text{first}$, with $\gamma_\text{first}$ being the angle corresponding to the first direction, and $\gamma_\text{max} = \dfrac{\pi}{n_\text{dir}}$. It is important to emphasise that the constraint relaxation, \cref{eq:norm_constraint_transformation}, is only applicable to 2-element vectors, while the norm constraint in Problem \ref{prob:original_formulation}, constraint \eqref{eq:umax_constraint_original}, is imposed on a 3-element vector, $\vec{f}^{r}_{k\vert k+1}$. The relaxation is, therefore,  applied the projection of the constraining sphere on each plane individually. The ROE dynamics (refer to \cite{di2018continuous}) possess unique characteristics that necessitates an accurate approximation of the constraining circle lying in the T-N plane, while coarse approximations of the constraints on the two other planes are acceptable. Concretely, the optimal solution to Problem \ref{prob:original_formulation} is expected to rarely incorporate radial thrust \cite{Gaias2015Impulsive, Belloni2023Relative} since it is known to be more expensive, from the Delta-V point of view, than relying solely on transversal thrust, especially when the reconfiguration can afford long maneuver times for large in-plane maneuvers. It is for this reason that the relaxation in \cref{eq:norm_constraint_transformation} is applied in the T-N plane with a larger number of directions, $n_\text{dir}$, than the number od directions used in the R-T and the R-N planes. Namely, we introduce $\bar{n}_\text{dir} = 4 < n_\text{dir}$ as the number of directions that approximate the constraining circles lying in the R-T and the R-N planes, with $\bar{\gamma}_{j} = \dfrac{2\parenth{j-1}\pi}{\bar{n}_\text{dir}} + \bar{\gamma}_\text{first} \; \forall j \in \bar{\dist{J}} = \left\{1,\; \hdots,\; \bar{n}_\text{dir}\right\}$, and with $\bar{\gamma}_\text{max} = \bar{\gamma}_\text{first}=\frac{\pi}{4}$. In this setting, the constraints in the R-T and the R-N planes are approximated by two rhombuses which cover only around $64\%$ of the original constraining circles.

Having introduced the constraint relaxations which transform the quadratic constraints into multiple affine ones, the reformulation of Problem \ref{prob:original_formulation} as a QP problem can be written as,
\begin{Problem}\label{prob:QP_formulation}
\begin{align}
& \min_{\mat{Y}, \mat{F}} \quad \textnormal{tr}\parenth{\mat{F}^{\intercal}\mat{F}} \qquad \text{subject to,} \nonumber\\
& \vec{y}_{0} = \vec{y}_{0}, \qquad \vec{y}_{m+1} = \vec{y}_{f}, \qquad \vec{y}_{k+1} =  \mat{\Phi}_{k\vert k+1} \vec{y}_{k} + \frac{a_{c}}{M} \mat{\Psi}_{k\vert k+1} \vec{f}^{r}_{k\vert k+1} \quad \forall k \in \dist{K},\\
& \vec{f}^{r}_{k\vert k+1} = \vec{0} \quad \forall k \in \dist{K}_{n}, \label{eq:u0_constraint_QP}\\
&\begin{bmatrix} 0 & \cos{\parenth{\gamma_{j}}} & \sin{\parenth{\gamma_{j}}} \end{bmatrix} \vec{f}^{r}_{k\vert k+1} \leq f_\text{max} \cos{\parenth{\gamma_\text{max}}},
\quad \forall j \in \dist{J} \; \& \; \forall k \in \dist{K}_{f}, \label{eq:umax_constraint_QP}\\
& \begin{bmatrix} \cos{\parenth{\bar{\gamma}_{j}}} & \sin{\parenth{\bar{\gamma}_{j}}} & 0\\
\cos{\parenth{\bar{\gamma}_{j}}} & 0 & \sin{\parenth{\bar{\gamma}_{j}}}  \end{bmatrix} \vec{f}^{r}_{k\vert k+1} \leq f_\text{max} \cos{\parenth{\bar{\gamma}_\text{max}}}, 
\quad \forall j \in \bar{\dist{J}} \; \& \; \forall k \in \dist{K}_{f}, \label{eq:umax_constraint_QP_R}
\end{align}
\end{Problem}

\begin{figure}[ht]
    \centering
    \includegraphics[width=\linewidth]{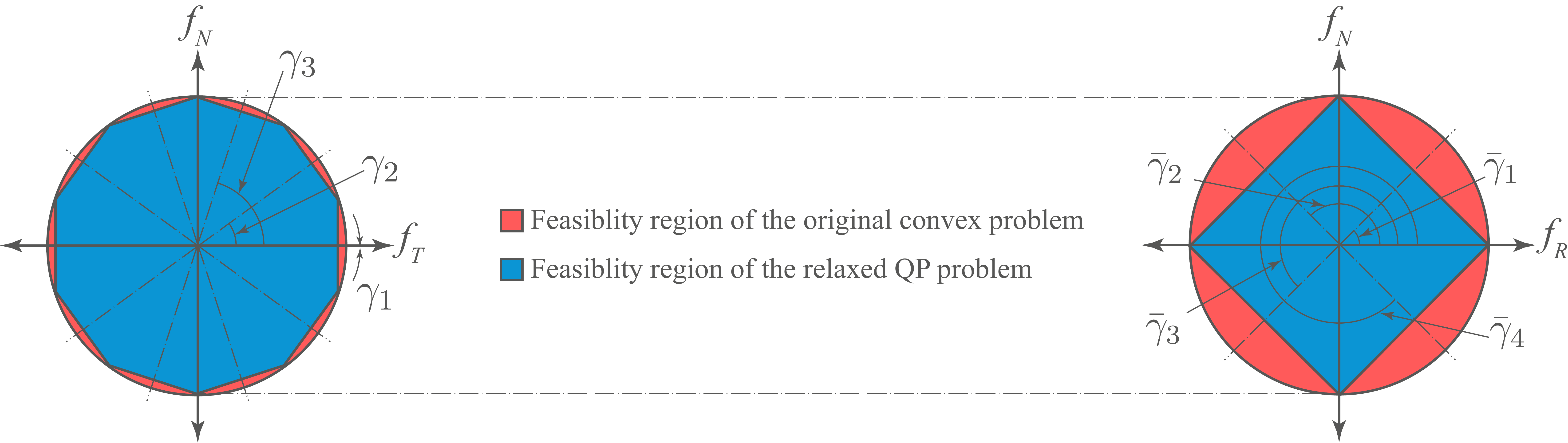}
    \caption{Feasibility region comparison between the original COP (Problem \ref{prob:original_formulation}) and the QP problem (Problem \ref{prob:QP_formulation})}
    \label{fig:feasibility_region_comparison_COP_and_QP}
\end{figure}

A graphical representation of the feasibility regions of the control thrust components is given in \cref{fig:feasibility_region_comparison_COP_and_QP} for both problems, Problem \ref{prob:original_formulation} and Problem \ref{prob:QP_formulation}. In \cref{fig:feasibility_region_comparison_COP_and_QP}, the constraint relaxation in \cref{eq:norm_constraint_transformation} is depicted for $n_\text{dir}=10$ and $\gamma_\text{first} = 0$ which covers approximately $94\%$ of the original constraining circle in the T-N plane.

It is important to note that while the relaxed constraint \eqref{eq:umax_constraint_QP_R} does not explicitly constrain the radial thrust component any better than the original circular constraint, it does affect the choice of the radial-transversal and normal-radial combinations.

\subsection{Parameters sensitivity analysis}
Investigating the two proposed formulations of the guidance problems, \ref{prob:original_formulation} and \ref{prob:QP_formulation}, and their graphical representation \cref{fig:Low-thrust-guidance-scheme}, it can be deduced that the parameters that are provided by the user and might need tuning are namely the time instances at which the thruster is switched on and off (or alternatively the time periods $T_{f,l}$ and $T_{n,l} \; \forall l \in \dist{L}$ as well as the allocated maneuver time, $t_{f}-t_{0}$). The choice of the time periods $T_{f,l}$ and $T_{n,l}$ is generally subject to mission time constraints, e.g. not being able to provide thrust during eclipse, during ground contact, or during scientific experiments. If no such mission constraints are present, fixing the time periods $T_{f,l}$ and $T_{n,l}$ comes as a natural choice, i.e. $T_{f,l} = T_{f},\; T_{n, l} = T_{n} \forall l \in \dist{L}$. For this specific case, a sensitivity analysis is performed to assess the feasibility of the optimization problem when the values of $T_{f}$ and $T_{n}$ change. In this sensitivity analysis, Problem \ref{prob:QP_formulation} was solved for $100$ randomly chosen initial conditions, i.e. $\vec{y}_{0}$, for the $91$ $T_{f}$-$T_{n}$ combination drawn from $T_{n} \in 
 \left\{60,\; 90,\; \hdots,\; 240\right\}\; \text{s}$ and $T_{f} \in \left\{0.025,\; 0.05,\; 0.1, \hdots, 0.5, 0.6, 0.7 \right\}\; \text{orbits}$.
Furthermore, since the difference between the initial and the final ROE vectors is what characterizes the maneuver, and not the values of the vectors themselves, only the value of $\vec{y}_{0}$ is chosen randomly, while the value of $\vec{y}_{f}$ is set to zero for the $9100$ experiments. Namely, the entries of the initial dimensional ROE vector, $\vec{y}_{0}$, are chosen randomly from the $\left[ -1 \; 1 \right] \; \text{km}$ range, except for the initial relative mean argument of longitude, $a\delta \lambda_{0}$, which is chosen randomly from the $\left[ -100 \; 100 \right] \; \text{km}$. The allocated time for the maneuver is fixed to $15$ orbits.\\

\begin{figure}[ht]
\begin{subfigure}[l]{0.5\linewidth}
    \centering
    \includegraphics[width=0.9\linewidth]{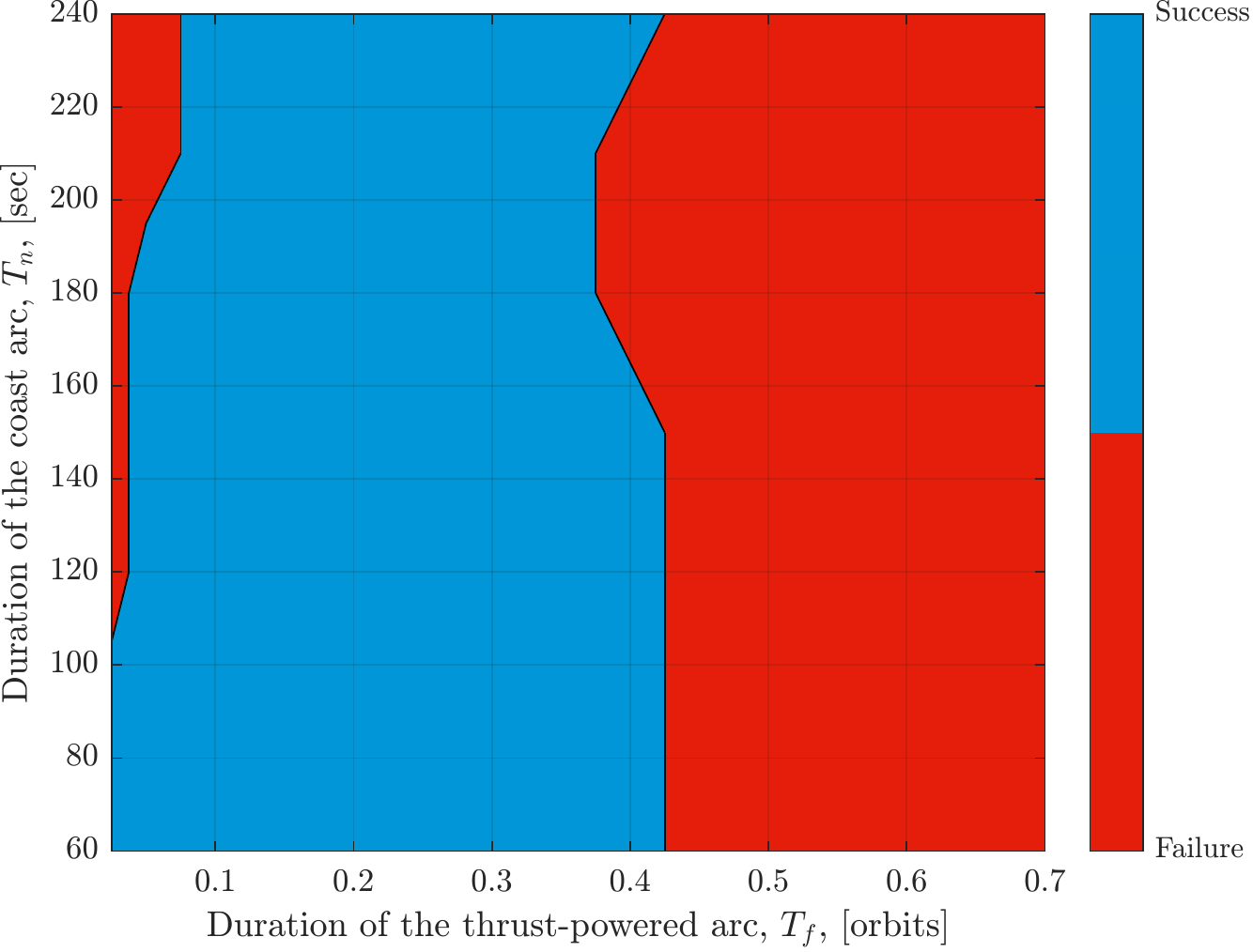}
    \caption{Optimization success region under varying $T_{f}$ and $T_{n}$} \label{fig:Sensitivity_analysis_success_failure_regions}
\end{subfigure}
\hfill
\begin{subfigure}[l]{0.5\linewidth}
    \centering
    \includegraphics[width=0.9\linewidth]{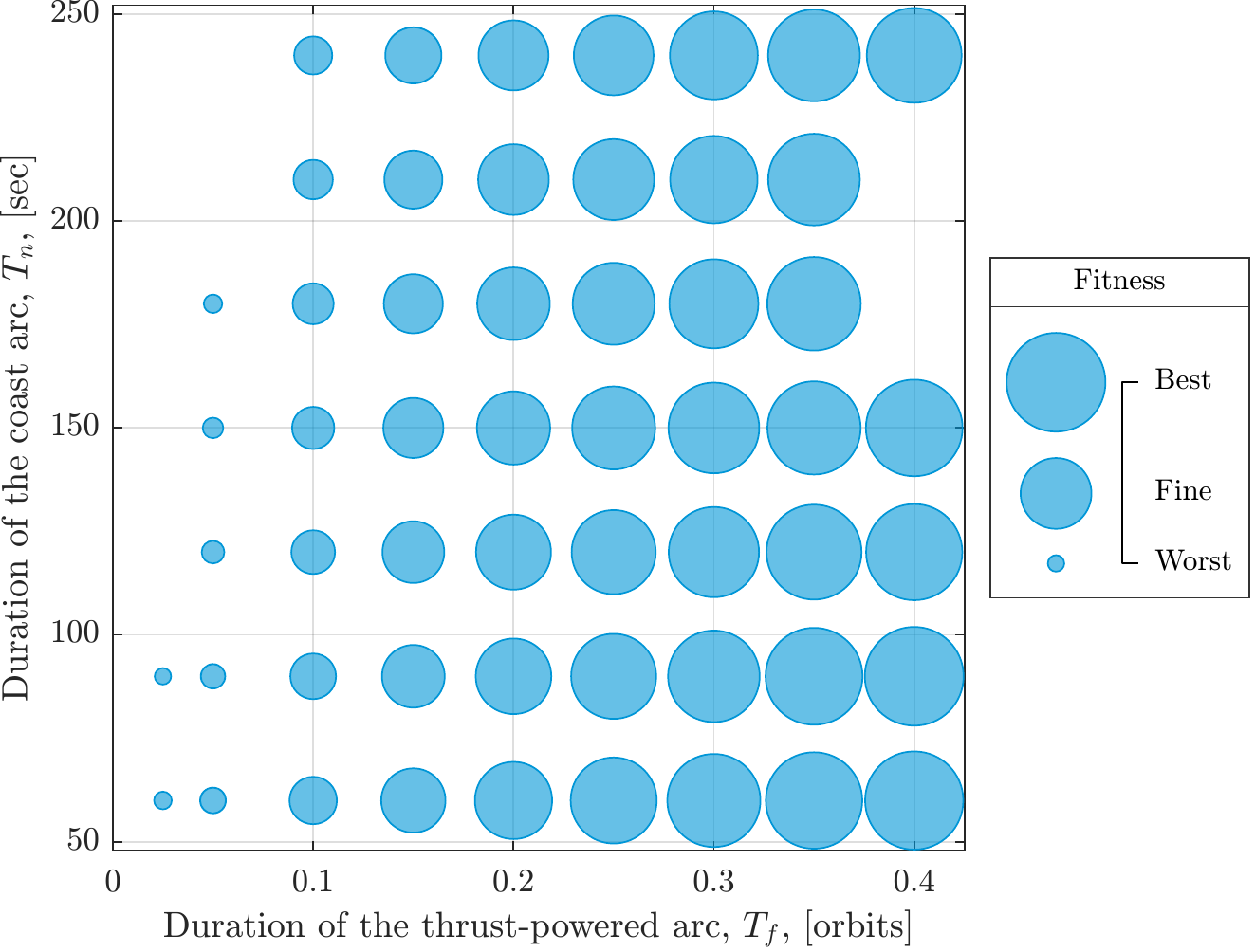}
    \caption{Fitness within the success region}
    \label{fig:Sensitivity_analysis_Tc_Tn_fitness}
\end{subfigure}
\caption{Sensitivity of the system to $T_{f}$ and $T_{n}$ variations}
\end{figure}

The output of the sensitivity analysis is the success and failure regions in the $T_{f}$-$T_{n}$ plane, where the success region is that in which the optimizer succeeded to find a feasible solution for all the $100$ random initial conditions. The success and failure regions of the aforementioned sensitivity study are depicted in \cref{fig:Sensitivity_analysis_success_failure_regions}.

Indeed, not all the points within the success region are the same from the fuel-efficiency point of view, since changing $T_{f}$ and $T_{n}$ does, in-turn, change the cost function. A fitness function is introduced to assess the competence of each $T_{f}$-$T_{n}$ combination within the success region. The adopted fitness function is the reciprocal of the average optimal cost over the $100$ initial conditions at each $T_{f}$-$T_{n}$ point.
The fitness
is calculated at each point in the success region and the results are presented in \cref{fig:Sensitivity_analysis_Tc_Tn_fitness}.
It is clear from \cref{fig:Sensitivity_analysis_Tc_Tn_fitness} that changing the value of $T_{n}$ barely changes the fitness for a fixed $T_{f}$ value, while it is obvious that the larger the value of $T_{f}$, the fitter the $T_{f}$-$T_{n}$ combination. Nevertheless, there comes a point where increasing $T_{f}$ drives the combination out of the success region (see \cref{fig:Sensitivity_analysis_success_failure_regions}). It is for this reason that the adopted value of $T_{f}$ in many of the numerical experiments to follow is set to $0.3$, which is a value that guarantees acceptable fitness, and is, at the same time, far away from the failure region.\\

Although the purpose of the sensitivity analysis is ultimately to help choosing adroit values for $T_{f}$ and $T_{n}$, since the time necessary for a slew maneuver heavily depends on the maximum slew rate, $\omega_\text{max}$, for any specific satellite, the value of $T_{n}$ can be calculated analytically according to the following formula,
\begin{equation}\label{eq:Tn}
    T_{n} = \dfrac{\pi}{\omega_\text{max}}+T_\text{safety}.
\end{equation}

The rationale behind \cref{eq:Tn} is that, since the maximum possible slew angle is $\pi$, the longest time period it takes the satellite to perform any slew, using the maximum angular speed, is $\frac{\pi}{\omega_\text{max}}$. Indeed, the satellite never uses the maximum angular speed throughout the whole slew, that is why the $T_\text{safety}$ term is added to make sure that the allocated time for the slew maneuver, $T_{n}$, is always more than enough. The importance of $T_\text{safety}$ is not limited to insuring the coast arc is sufficient for the slew maneuver, it also proves to be useful in closing the loop as will be discussed in \cref{sec:Closed_loop}. For Triton-X, the maximum slew rate is $2^{\circ}/\text{s}$, and hence $T_{n}$ is calculated to be $100$ s after choosing $T_\text{safety}=10$ s.\\

\section{Guidance scheme validation}\label{sec:Guidance_validation}
After comparing the performance of many different convex QP solvers, the free open-source Operator Splitting Quadratic Program (OSQP) \cite{osqp} stood as the fastest, and was hence chosen to solve Problem \ref{prob:QP_formulation} in the coming discussions. 
In order to validate the proposed guidance scheme for the targeted maneuver spectrum, the optimization problem is solved for a variety of maneuvers, large and otherwise, using the parameters chosen based on the sensitivity study, and the results of one of these tests is reported here. In that experiment, the initial and final dimensional ROE vectors are randomly chosen to be, $\vec{y}_{0} = \left[-55.6 \;  7414.7 \;  -58.7 \;  83.7 \;  -2.3 \;  22.4 \right]^{\intercal} \; \text{m}$ and $\vec{y}_{f} = \vec{0}$, while the initial orbit of the chief is defined in terms of the osculating orbital elements such that $\tilde{\vec{\alpha}}_{c, 0} = \left[7121\;\text{km} \; 0^{\circ} \; 10^{-5} \; 0 \; 45^{\circ} \; 0^{\circ} \right]^{\intercal}$.
The full parameters list of the reported maneuver is presented in \cref{tab:guidance_validation_simulation_parameters}. Note that choosing $\pmb{y}_{f}$ to be zeros means that the two satellites are required to rendezvous at the final time.
{\renewcommand{\arraystretch}{\arraystretchfortable}
\begin{table}[ht]
    \centering
    \caption{Parameters used in the guidance scheme validation simulation}     \begin{tabular}{cccccccc}
        \hline
        \hline
        {$t_{f}-t_{0}$ \; [\text{orbits}]} & {$T_{f}$ \; [\text{orbits}]} & {$T_{n} \; [\text{s}]$} & {$n_\text{dir} \; [\text{-}]$} & $\gamma_\text{first} \; [^{\circ}]$ & $M \; [\text{kg}]$ & {$f_\text{max} \; [\text{mN}]$} &  {$\omega_\text{max} \; [^{\circ}/\text{s}]$}\\
        \hline
        $5$ & $0.3$ & $100$ & 
        $12$ & $0$ & $200$ & $7$ & {$2$}\\
        \hline
        \hline
    \end{tabular}
\label{tab:guidance_validation_simulation_parameters}
\end{table}
}
It is to be noted that the satellite mass, the maximum thrust, and the maximum slew rate are extracted from the publicly available Triton-X specifications\footnote{The Triton-X brochure can be found at \href{https://luxspace.lu/resources/}{https://luxspace.lu/resources/}.}. Moreover, $n_\text{dir}$ is set to $12$ as this is the least number of directions that approximates the  constraining circle by multiple affine constraints while covering at least $95\%$ of the its area.\\

\begin{figure}[ht]
    \centering
    \includegraphics[width=\linewidth]{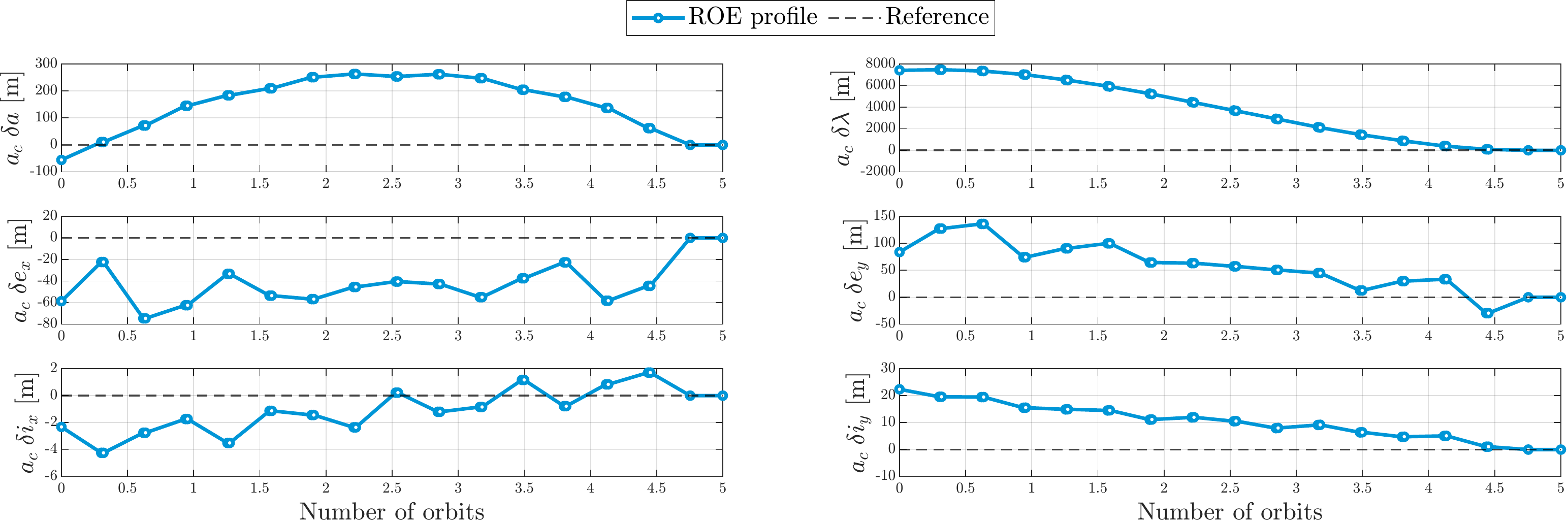}
    \caption{Optimized dimensional ROE profile}
    \label{fig:Guidance_ROE_profile}
\end{figure}

In \cref{fig:Guidance_ROE_profile}, the dimensional ROE profile, optimized by the guidance scheme, is depicted. The plots in \cref{fig:Guidance_ROE_profile} reveal that the main goal of the guidance algorithm, which is to rendezvous with the virtual chief at the final time, is achieved. Moreover, the $\delta \lambda$ error is corrected through building $\delta a$ momentum instead of firing in the radial direction. 

\begin{figure}[ht]
\begin{subfigure}[l]{0.5\linewidth}
    \centering
    \includegraphics[width=0.9\linewidth]{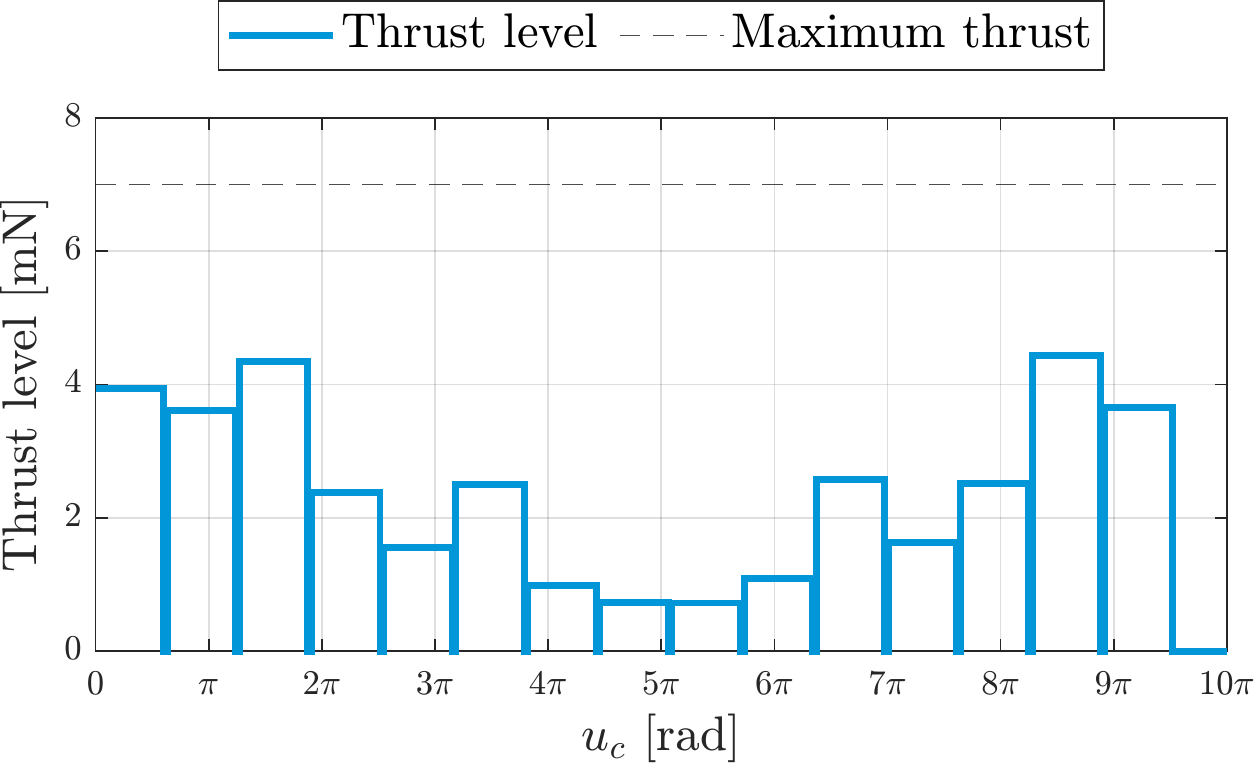}
    \caption{Thrust norm}
    \label{fig:Guidance_thrust_norm_profile}
\end{subfigure}
\hfill
\begin{subfigure}[l]{0.5\linewidth}
    \centering
    \includegraphics[width=0.9\linewidth]{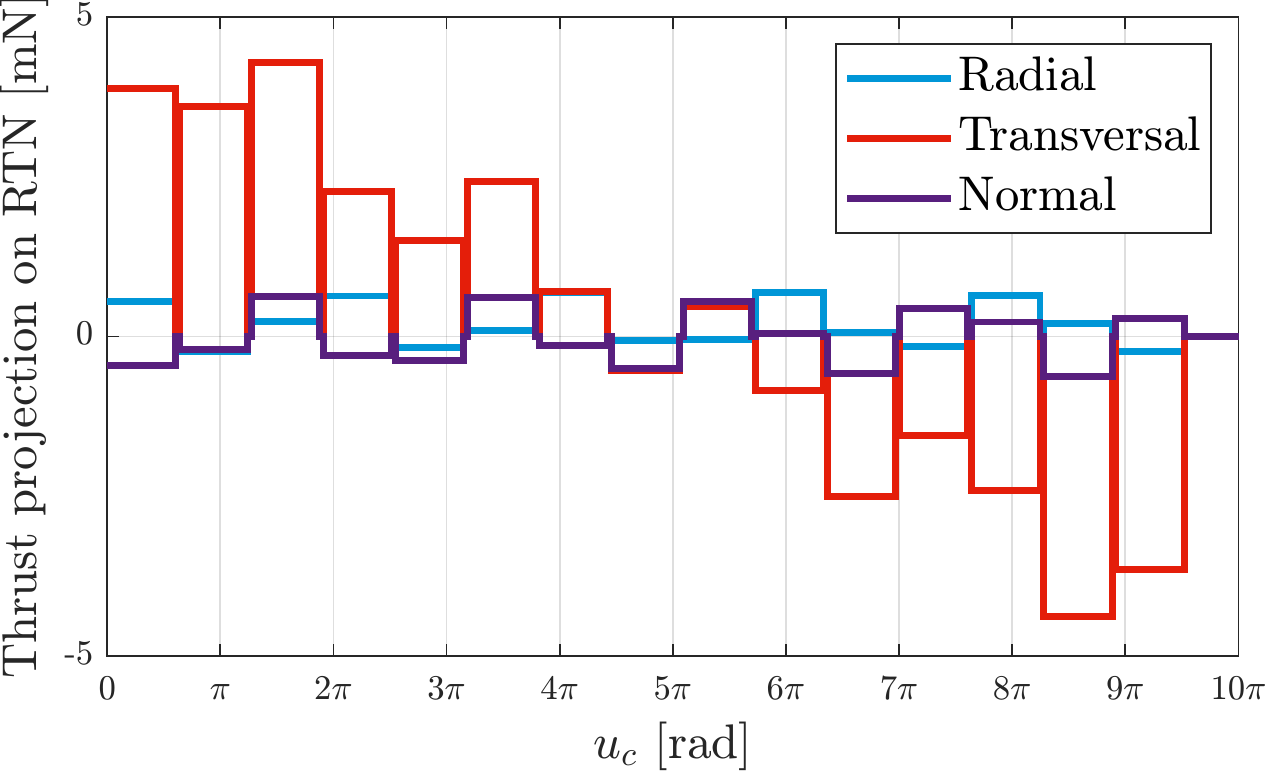}
    \caption{Thrust projection into the RTN frame}
    \label{fig:Guidance_thrust_profile}
\end{subfigure}
\caption{Optimized thrust profile}
\end{figure}

The level of thrust provided by the onboard electric propulsion system is presented in \cref{fig:Guidance_thrust_norm_profile}. It is clear that the maximum thrust constraint is respected thanks to the relaxations \eqref{eq:umax_constraint_QP} and \eqref{eq:umax_constraint_QP_R} that approximate the original quadratic constraint \eqref{eq:umax_constraint_original}.
Furthermore, the fact that the satellite uses minimal radial thrust, as expected, is evident in \cref{fig:Guidance_thrust_profile} which depicts the projection of the thrust into the RTN frame.  

To insure that the optimized trajectory is compliant with the maximum slew rate constraint, the mean angular rate is calculated at each time step and is shown in \cref{fig:Guidance_angle_rate_profile}. 
It is obvious that the mean required angular speed is conceivably less than $\omega_\text{max}$ since the allocated time for the attitude maneuver, $T_{n}$, is always more than the time required for the most extreme attitude maneuver.

Satellite missions usually have constraints on the timing of the orbital maneuvers, which arise from not being able to perform any thruster firing for example during ground contact or during eclipse periods.
Unlike impulsive-thrust absolute/relative orbit correction maneuvers, low-thrust maneuvers may require too long to execute, days in some cases. It is for this reason that low-thrust guidance algorithms need to accommodate the no-thrust periods during the orbital maneuver itself. 
One of the main contributions of this note is the ability of the proposed guidance scheme to adapt to different scenarios where the thruster is required to shut down for known extended periods without the need to change the structure of the problem. The duration of each coast arc, $T_{n, l}$ (see \cref{fig:Low-thrust-guidance-scheme}), which is a user-input, is the only thing that needs to be adapted according to the operational constraints.
To validate the ability of the trajectory optimization routine to accommodate long coast arcs, two different scenarios are introduced where the maneuver duration as well as the initial and final ROE vectors as fixed for both scenarios and only the no-thrust periods are varied. The thruster off-periods are arbitrarily chosen for each of the two scenarios and are reported in \cref{tab:guidance_with_operational_constraints}. 
The initial and final dimensional ROE vectors for both scenarios are set to, $\vec{y}_{0} = \left[187 \;  945 \; 189 \;  86 \; 79 \;  -114 \right]^{\intercal} \; \text{m}$ and $\vec{y}_{f} = \left[0 \;  412 \;  389 \;  -96 \;  153 \;  -198 \right]^{\intercal} \; \text{m}$, and the simulation parameters are identical to those reported in \cref{tab:guidance_validation_simulation_parameters} except for $T_{f}$ which is set to $0.1$ orbits.
{\renewcommand{\arraystretch}{\arraystretchfortable}
\begin{figure}[ht]
    \begin{minipage}{0.5\textwidth}    
    \centering
    \caption{Adopted no-thrust periods for two simulation scenarios}
    \begin{tabular}{ccc}
        \hline
        \hline
        ~ & No-thrust periods [orbits]\\      
        \hline
        {Scenario 1} & {$\left\{0.5\text{-}1, \;1.5\text{-}2, \;2.5\text{-}3, \;3.5\text{-}4, \;4.5\text{-}5\right\}$}\\
        {Scenario 2} & {$\left\{
        0.25\text{-}0.5,\; 1.25\text{-}2.25,\; 3\text{-}3.25,\; 4.25\text{-}4.75\right\}$}\\
        \hline
        \hline
    \end{tabular}
    \label{tab:guidance_with_operational_constraints}
    \end{minipage}
    \hfill
    \begin{minipage}{0.45\textwidth}
        \centering
        \includegraphics[width=\linewidth]{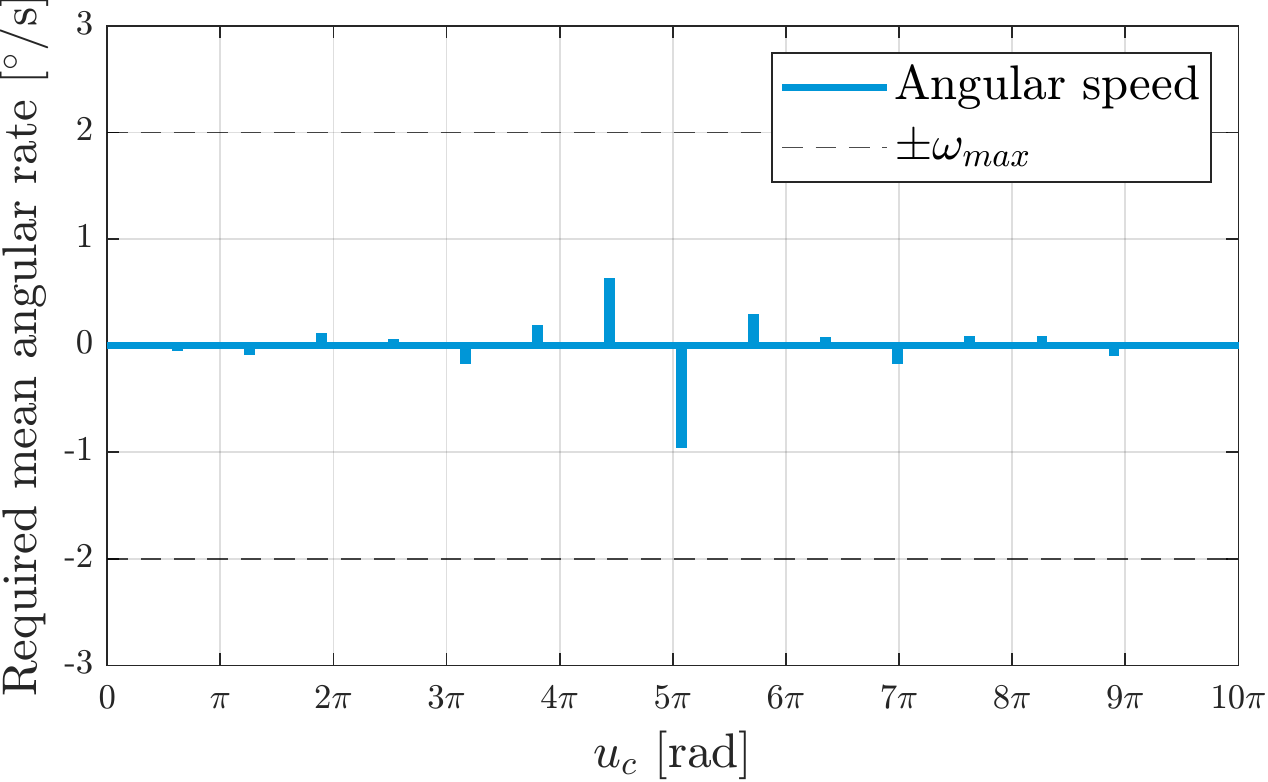}
        \caption{Optimized angular profile}
        \label{fig:Guidance_angle_rate_profile}
    \end{minipage}
\end{figure}
}

\begin{figure}[ht]
    \centering
    \includegraphics[width=\linewidth]{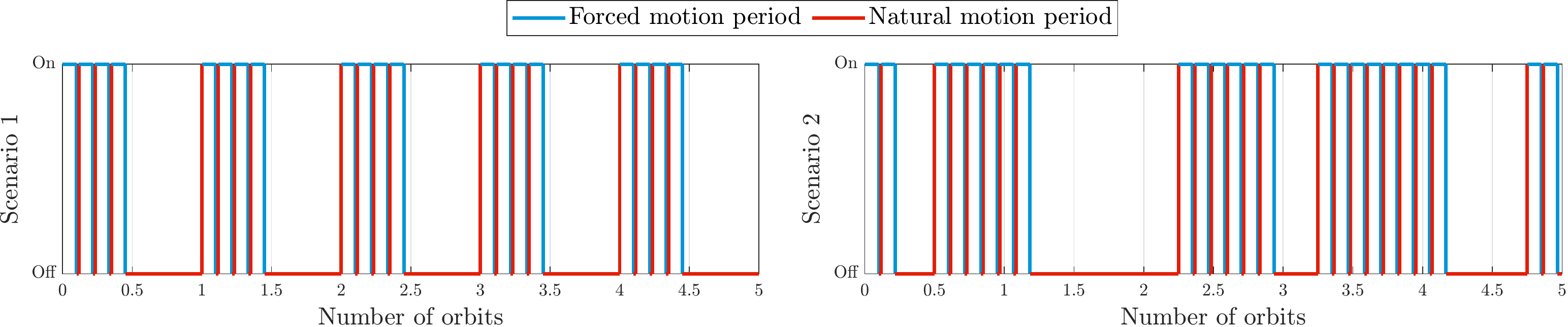}
    \caption{Expected behaviour of the thruster in the presence of the thruster off-periods in \cref{tab:guidance_with_operational_constraints}}
    \label{fig:Guidance_time_vector_operational_Constraints}
\end{figure}

The user-defined thruster no-thrust intervals for the two defined scenarios can be seen graphically in \cref{fig:Guidance_time_vector_operational_Constraints} where the expected behaviour of the thruster is presented. 
It is important to emphasise that while the no-thrust periods are seen to appear with a regular pattern (from the halfway point to the end of each orbit) for scenario 1, this need not to be the case for operational time constraints, which is reflected in scenario 2.\\

\begin{figure*}[ht]
    \centering
    \includegraphics[width=\linewidth]{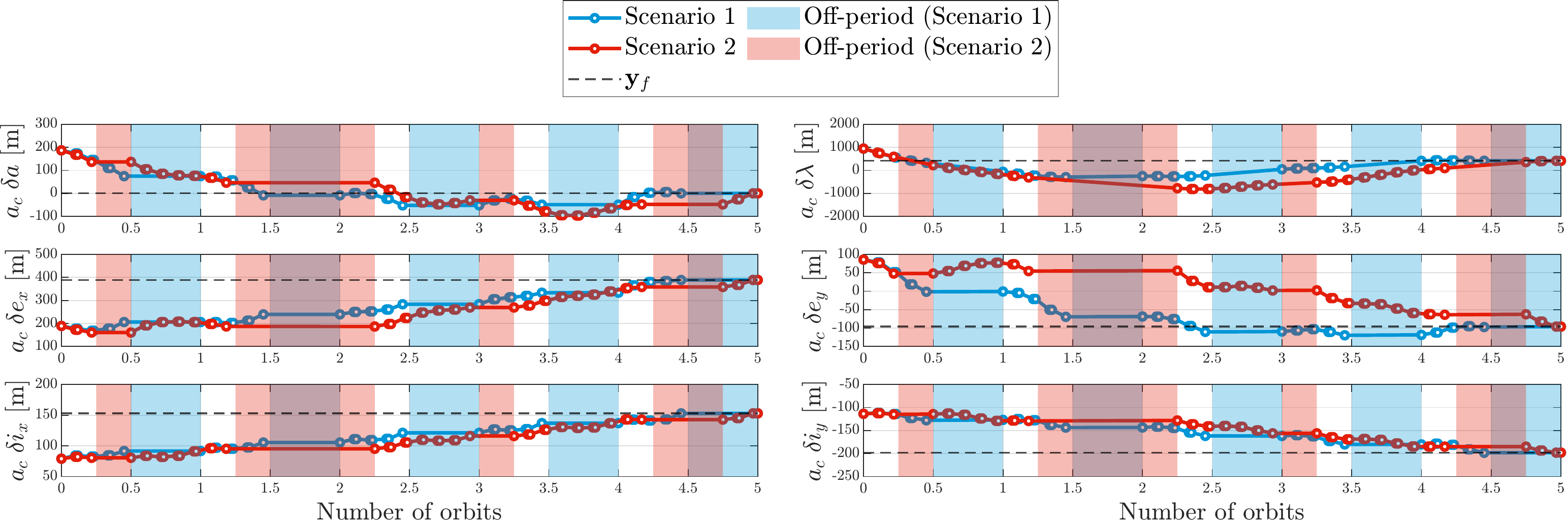}
    \caption{Guidance dimensional ROE profile  under operational constraints}
    \label{fig:Guidance_ROE_profile_operational_Constraints}
\end{figure*}

The dimensional ROE profile, predicted by the guidance algorithm, is depicted in \cref{fig:Guidance_ROE_profile_operational_Constraints} for both scenarios, where the thruster off-periods are also shown. It is obvious that the $\delta \lambda$ signal is conceivably evolving even when no thrust is provided, since it can be manipulated not only directly through input thrust, but also indirectly through the non-zero value of $\delta a$. Notably, the total Delta-V is different for each of the scenarios, as it counts to $0.286 \; \text{m}/\text{s}$ in the first scenario and to  $0.303 \; \text{m}/\text{s}$ in the second. 

\section{Closing the loop}\label{sec:Closed_loop}
\begin{figure*}[t]
    \centering
    \includegraphics[width=\linewidth]{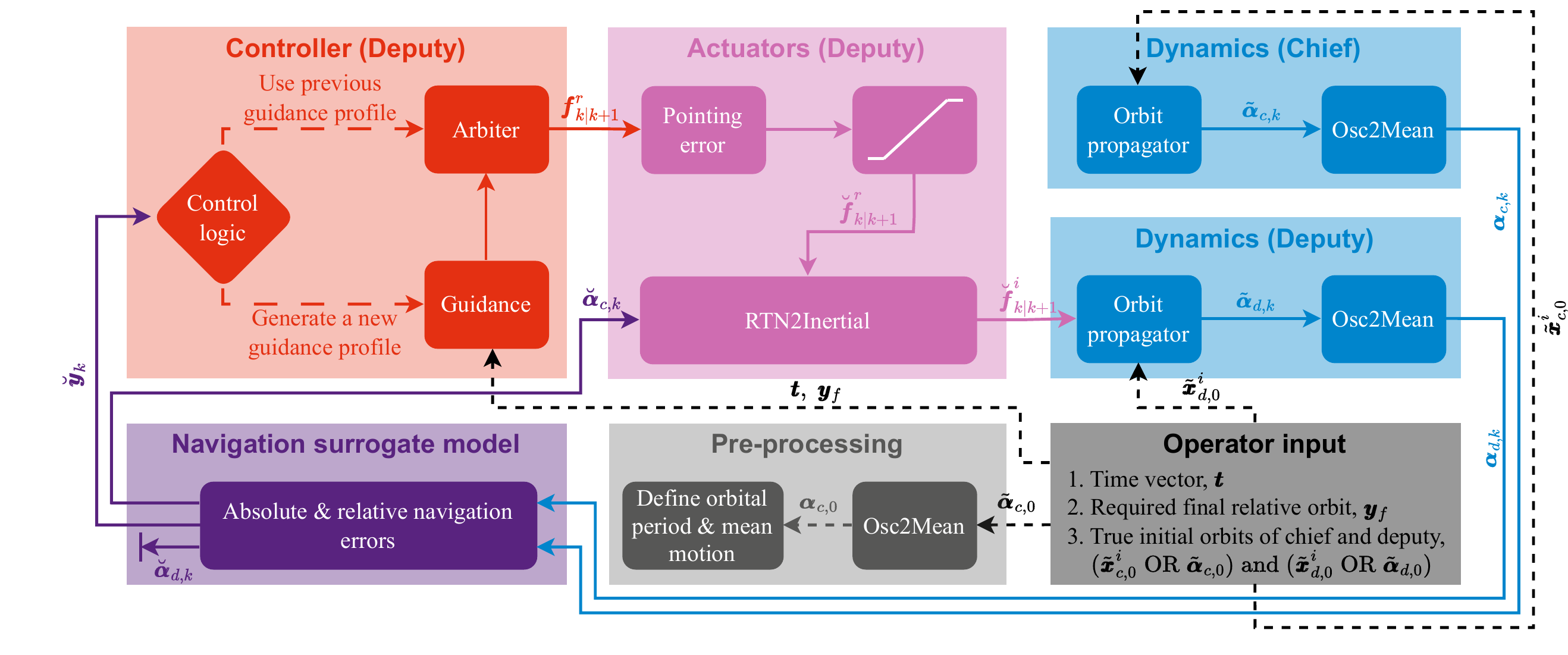}
    \caption{Deputy's module execution logic used in numerical simulations}
    \label{fig:Closed-loop-relative-simulation}
\end{figure*}

The proposed trajectory optimization scheme is an open-loop control system that cannot handle disturbances or model inaccuracies. It is for this reason that the loop has to be closed to react to system changes in real-time. In this section, the feedback control loop is studied, and the exact location of the guidance module within the closed loop is discussed.
The simulated module execution logic onboard of the deputy is depicted in \cref{fig:Closed-loop-relative-simulation}, where the solid arrows signify the main signals that are recurrently passed, and the dashed lines are those which are passed only once before the beginning of the maneuver.
In \cref{fig:Closed-loop-relative-simulation}, "Osc2Mean" is the function that transforms osculating orbital elements to mean ones \cite{Gaias2020Analytical}, and "RTN2Inertial" is the method that rotates any vector from $\set{F}^{r}$ to $\set{F}^{i}$ given the position and velocity of the chief (or alternatively its orbital elements) at the time instant in question.
Moreover, the breve accent, $\breve{\parenth{\cdot}}$, signifies a quantity which is disturbed by either one or a combination of a) Estimation errors, e.g. $\breve{\vec{\alpha}}_{d, k}$; b) ADCS inaccuracies, e.g. $\breve{\vec{f}}^{i}_{k\vert k+1}$; or c) Physical constraints of the actuators, e.g. $\breve{\vec{f}}^{i}_{k\vert k+1}$.
Note that in quantities with double subscripts, the first subscript refers to the spacecraft, either chief or deputy, and the second refers to the time instant at which the quantity is evaluated, e.g. $\breve{\vec{\alpha}}_{d, k}$ is the perturbed orbital elements vector of the deputy at time $t_{k}$.\\

Since the navigation module is out of the scope of this note, and in order to take the hardware inaccuracies and physical limitations into account within the numerical simulations, surrogate models for estimation inaccuracies as well as for the physical limitations and ADCS errors are introduced.\\
The physical limitations are only present in the saturation block in \cref{fig:Closed-loop-relative-simulation}, which can be easily implemented in the numerical simulations. In fact, this saturation is taken into account in the guidance implementation as the maximum thrust constraint (see constraints \eqref{eq:umax_constraint_QP} and \eqref{eq:umax_constraint_QP_R} of Problem \ref{prob:QP_formulation}), nonetheless, it is also implemented in the control loop as a safeguard. The navigation surrogate model uses the input mean orbital elements of the deputy and the chief as ground truth. The following procedure is used on the chief spacecraft,
\begin{equation}\label{eq:navigation_surrogate}
    \vec{x}_{c,k}^{i} =  \text{OE2Cart}(\vec{\alpha}_{c,k}),\qquad
    \breve{{\vec{x}}}_{c,k}^{i} = {\vec{x}}_{c,k}^{i} + \dist{N}\parenth{\vec{0},\; \mat{\Sigma}_{\vec{x}_{c}}},\qquad
    \breve{\vec{\alpha}}_{c,k} = \text{Cart2OE}(\breve{\vec{x}}_{c,k}^{i}),
\end{equation}
where the two methods "OE2Cart" and "Cart2OE" are those which transform the Orbital elements vector into Cartesian state vector and vice versa, and $\dist{N}\parenth{\mu,\;\sigma^{2}}$ is a normally distributed random variable with $\mu$ as its mean and $\sigma^{2}$ as its variance. Hence, $\mat{\Sigma}_{{\vec{x}}_{c}}$ is the covariance matrix of the random noise affecting the estimation of the Cartesian state of the chief satellite, which is defined as,
\begin{equation}
    \mat{\Sigma}_{{\vec{x}}_{c}} = \text{diag}\parenth{\sigma_{\vec{r}_{c}}^{2},\; \sigma_{\vec{r}_{c}}^{2},\; \sigma_{\vec{r}_{c}}^{2},\; \sigma_{\vec{v}_{c}}^{2},\; \sigma_{\vec{v}_{c}}^{2},\; \sigma_{\vec{v}_{c}}^{2}}
\end{equation}
where $\text{diag}\parenth{\cdot, \cdot, \hdots}$ is a function that creates a diagonal matrix, with zero off-diagonal elements, from its input arguments, $\sigma_{\vec{r}_{c}}^{2}$ and $\sigma_{\vec{v}_{c}}^{2}$ are the variances of the 1-dimensional (in the x, the y or the z directions) position and velocity estimation errors respectively.
It is worth mentioning that the mean orbital elements of the deputy can be disturbed using the same model in \cref{eq:navigation_surrogate}, however using different variances for the 1-dimensional position and velocity estimation errors, namely, $\sigma_{\vec{r}_{d}}$ and $\sigma_{\vec{v}_{d}}$. Moreover, relative navigation is, in general, more accurate than the absolute one  \cite{DAmico2009navigation, Mahfouz2023GNSS-based}. Hence the surrogate model for the relative navigation needed to be more than just converting the estimated absolute orbital elements of the chief and the target to a ROE vector. The following surrogate model is used instead,
\begin{equation}\label{eq:y_navigation_surrogate}
    \delta\vec{\alpha} = \text{OE2ROE}\parenth{\vec{\alpha}_{c}\parenth{t},\; \vec{\alpha}_{d}},\qquad
    \breve{\vec{y}} = a_{c}\delta\vec{\alpha} + \dist{N}\parenth{\vec{0},\; \mat{\Sigma}_{\vec{y}}},
\end{equation}
where OE2ROE is the method that transforms the orbital elements of the chief and deputy to a ROE vector according to \cref{eq:ROE}, and $\mat{\Sigma}_{\vec{y}}$ is the covariance matrix of the zero mean normally distributed random disturbance vector, which can be expressed as,
\begin{equation}
    \mat{\Sigma}_{\vec{y}} = \text{diag}\parenth{\sigma_{\vec{y}}^{2},\; \sigma_{\vec{y}}^{2},\; \sigma_{\vec{y}}^{2},\; \sigma_{\vec{y}}^{2},\; \sigma_{\vec{y}}^{2},\; \sigma_{\vec{y}}^{2}}.
\end{equation}

Lastly, the "Pointing error" block in \cref{fig:Closed-loop-relative-simulation} is using the the following surrogate model,
\begin{equation}\label{eq:q_surrogate}
    \quat{q}_\text{pe} = \begin{bmatrix}
        \cos{\parenth{{\zeta_\text{pe}}/{2}}} & \sin{\parenth{{\zeta_\text{pe}}/{2}}} \hat{\vec{q}}_\text{pe}^{\intercal}
    \end{bmatrix}^{\intercal},\qquad
    \breve{\vec{f}}^{r}_{k\vert k+1} = \quat{q}_\text{pe} \circ \vec{f}^{r}_{k\vert k+1}  \circ  \tilde{\quat{q}}_\text{pe},
\end{equation}
where $\quat{q}_\text{pe}$ is the thruster misalignment unit quaternion, with $\tilde{\quat{q}}_\text{pe}$ being its quaternion conjugate, $\zeta_\text{pe}$ is the pointing error angle, which can be extracted from Triton-X brochure, $\hat{\vec{q}}_\text{pe}$ is a random 3-element unit vector, and $\circ$ is the quaternion multiplication operator.\\

The "Controller" block collection in \cref{fig:Closed-loop-relative-simulation} works much like a Model Predictive Control (MPC) where the prediction horizon spans from the current time to the user-defined maneuver end time, and the control and prediction horizons are identical. In this setting, the control profile optimization as well as the state prediction from the current time to the maneuver final time is done recurrently by the trajectory optimization scheme (Problem \ref{prob:QP_formulation}), and the function of the "Arbiter" block is only to choose the proper input thrust from the provided guidance profile, namely the first or the second thrust vector from the guidance control profile depending on whether the current time lies within a forced or a natural translational motion period. The logic of the closed loop which is used in the validation simulations is elaborated upon in Algorithm \ref{alg:Sim-closed-loop-logic}. In this algorithm, $\vec{t}$ is the user-defined time vector for the guidance (see \cref{fig:Low-thrust-guidance-scheme}), $\vec{y}_{f}$ is the required final dimensional ROE vector, and $\epsilon$ is a tunable threshold.
\begin{algorithm}
\setstretch{1.2}
\KwIn{$\vec{t}, \;\vec{y}_{f}, \; \epsilon,\; \tilde{\vec{\alpha}}_{c,0},\; \tilde{\vec{\alpha}}_{d,0}$}  

\For{$k= 0 \to m$}{
$t_{k}, \; t_{k+1} \gets$ Elements with indices $k \to \parenth{k+1}$ in $\vec{t}$\;
$\text{NoUpdateFlag} \gets \text{false}$\;
\eIf{$k=0$}{
    $\text{GuidanceFlag} \gets \text{true}; \hfill \vec{t}_{k} \gets \vec{t}; \hfill \vec{y}_\text{pred} \gets \vec{\infty}$\;
}{
    $t_{k-1}\gets$ Element with index $\parenth{k-1}$ in $\vec{t}$\;
    $\tilde{\vec{x}}_{c, k}, \; \tilde{\vec{\alpha}}_{c,k} \gets$ Propagate chief's orbit $t_{k-1} \to t_{k}$\;
    \eIf{$k$ \textnormal{is even}}{
    $\text{GuidanceFlag} \gets \text{true}$\;
    $\vec{t}_{k} \gets$ Elements $\parenth{k+2} \to \textbf{last}$ in $\vec{t}$\;
    $\vec{y}_\text{pred} \gets$ Column $2$ in $\mat{Y}_{k}$\;
    $\tilde{\vec{x}}_{d, k}, \; \tilde{\vec{\alpha}}_{d,k} \gets$ Propagate deputy $t_{k-1} \to t_{k}$\;
    }{
    $\text{GuidanceFlag} \gets \text{false}$\;
    }
}
$\vec{\alpha}_{c,k} \gets \text{Osc2Mean}\parenth{\tilde{\vec{\alpha}}_{c,k}}$\;

\eIf{$\textnormal{GuidanceFlag}$}{
    $\vec{\alpha}_{d,k} \gets \text{Osc2Mean}\parenth{\tilde{\vec{\alpha}}_{d,k}}$\;
    $\breve{\vec{x}}_{c, k}, \; \breve{\vec{\alpha}}_{c,k} \gets$ Apply \cref{eq:navigation_surrogate} on ${\vec{\alpha}}_{c,k}$\;
    $\breve{\vec{y}}_{k} \gets$ Apply \cref{eq:y_navigation_surrogate} on ${\vec{\alpha}}_{c,k}$ and ${\vec{\alpha}}_{d,k}$\;
    \eIf{$\left\Vert \breve{\vec{y}}_{k} - \vec{y}_\text{pred}\right\Vert \geq \epsilon $}{
    $\mat{Y}_{k}, \; \mat{F}_{k} \gets$ Solve Problem \ref{prob:QP_formulation} using $\vec{t}_{k}$, $\breve{\vec{y}}_{k}$, $\vec{y}_{f}$\;
    \If{$\textnormal{Solver did not succeed}$}{
    \If{$k=0$}{
    \ColoredComment{Problem is infeasible for the given $\vec{t}$ vector}
    $\textbf{break}$\;}
    $\text{NoUpdateFlag} \gets \text{true}$\;}
    }{
    $\text{NoUpdateFlag} \gets \text{true}$\;
    }
    \If{$\textnormal{NoUpdateFlag}$}{
    \ColoredComment{Sticking to the previous guidance profile}
    $\mat{Y}_{k} \gets$ Columns $2 \to$ $\textbf{last}$ in $\mat{Y}_{k-2}$\;
    $\mat{F}_{k} \gets$ Columns $2 \to$ $\textbf{last}$ in $\mat{F}_{k-2}$\;
    }
    $\vec{f}^{r}_{k\vert k+1} \gets$ Column $0$ in $\mat{F}_{k}$, divided by $\frac{a_{c}}{M}$\;
    }{
    $\vec{f}^{r}_{k\vert k+1} \gets$ \vec{0}\;
    }
    $\breve{\vec{f}}^{r}_{k\vert k+1} \gets$ Apply \cref{eq:q_surrogate} and saturation on ${\vec{f}}^{r}_{k\vert k+1}$\;
    $\breve{\vec{f}}^{i}_{k\vert k+1} \gets$ Rotate $\breve{\vec{f}}^{r}_{k\vert k+1}$ from $\set{F}^{r}$ to $\set{F}^{i}$ using $\breve{\vec{x}}_{c, k}$\;
    $\tilde{\vec{x}}_{d, k+1}, \; \tilde{\vec{\alpha}}_{d,k+1} \gets$ Propagate deputy $t_{k} \to t_{k+1}$ with constant thrust $\breve{\vec{f}}^{i}_{k\vert k+1}$\;
}
\caption{Simulated control loop}
\label{alg:Sim-closed-loop-logic}
\end{algorithm}

The first guidance profile is conceivably calculated before the maneuver starting time, $t_{0}$. As suggested by Algorithm \ref{alg:Sim-closed-loop-logic}, the trajectory might need to be optimized during the execution of the maneuver, and this is exactly when the importance of including $T_\text{safety}$ within $T_{n}$ becomes apparent. While $T_\text{safety}$ guarantees that $T_{n}$ is sufficient for the most stringent attitude redirection maneuver as \cref{eq:Tn} suggests, it is also used to optimize the next guidance profile if one needs to be optimized, for example before $t_{2}$ or $t_{4}$ (see \cref{fig:Low-thrust-guidance-scheme}), which is very valuable from the practical point of view.\\

To test the performance of the closed loop system using the proposed guidance scheme, it has been benchmarked against the MPC proposed in \cite{Belloni2023Relative}. The Out-Of-Plane (OOP) maneuver in \cite{Belloni2023Relative} is chosen for the benchmark because such maneuvers can give a clear insight about the fuel-optimality ($\equiv \; \Delta V$-optimality for uni-directional propulsion systems) of the control algorithm since only thrust in the normal direction is required. In fact, the exact locations (in terms of $u_{c}$) in which impulsive thrust should be provided for a $\Delta V$-optimal OOP maneuver can be calculated analytically \cite{Gaias2015Impulsive}, which makes it easy for a human eye to recognise $\Delta V$-optimal thrust profiles. For the low-thrust case, the thrust in the normal direction is expected to be distributed evenly around these locations for the control profile to be close to $\Delta V$-optimality, while the thrust components in the radial and transversal directions are expected to be around zero. The initial and final dimensional ROE vectors for benchmark maneuver as well as the chief's initial orbit are drawn from \cite{Belloni2023Relative} to be $\vec{y}_{0} = \left[0 \;  0 \;  273 \;  0 \;  10 \;  70 \right]^{\intercal} \; \text{m}$, $\vec{y}_{f} = \left[0 \;  0 \; 273 \;  0 \;  400 \;  120 \right]^{\intercal} \; \text{m}$, and $\tilde{\vec{\alpha}}_{c, 0} = \left[ 6828\;\text{km} \; 0^{\circ} \; 10^{-5} \; 0 \; 78^{\circ} \; 0^{\circ} \right]^{\intercal}$.
A summary of the parameters used in the benchmarck maneuver is provided in \cref{tab:benchmarck_simulation_parameters}. All the parameters that are not reported are identical to the ones in \cref{tab:guidance_validation_simulation_parameters} except for the length of the forced translational motion periods, which is set to $T_{f} = 0.3\; \text{orbits}$ for the first $6$ orbits of the maneuver and to $T_{f} = 0.1\; \text{orbits}$ for the final orbit to allow for a more precise approach to the target relative orbit. 
{\renewcommand{\arraystretch}{\arraystretchfortable}
\begin{table}[ht]
    \centering
    \caption{Benchmark simulation parameters}
    \begin{tabular}{cccccc}
        \hline
        \hline
        {$t_{f}-t_{0}$ \; [\text{orbits}]} & {$\sigma_{\vec{r}_{d}} \equiv \sigma_{\vec{r}_{c}} \; [\text{m}]$} & {$\sigma_{\vec{v}_{d}} \equiv \sigma_{\vec{v}_{c}} \; [\text{m}/\text{s}]$} & {$\sigma_{\vec{y}} \; [\text{m}/\text{s}]$} & 
        {$\zeta_\text{pe} \; [\text{arcs}]$} & {$\epsilon \; [\text{m}]$}\\
        \hline
        {$7$} & {$10$} & {$0.5$} & {$1$} & 
        {$25$} & {$5$} \\
        \hline
        \hline
    \end{tabular}
    \label{tab:benchmarck_simulation_parameters}
\end{table}
}
It is important to note that while $t_{f}-t_{0}$, $T_{f}$, $T_{n}$, and $\epsilon$ are tunable parameters, the values of $\sigma_{\vec{r}_{d}}$, $\sigma_{\vec{v}_{d}}$ and $\zeta_\text{pe}$ are extracted from the specification sheets of Triton-X even if the benchmark maneuver is assuming a different satellite.\\

\begin{figure}[ht]
    \centering
    \includegraphics[width=\linewidth]{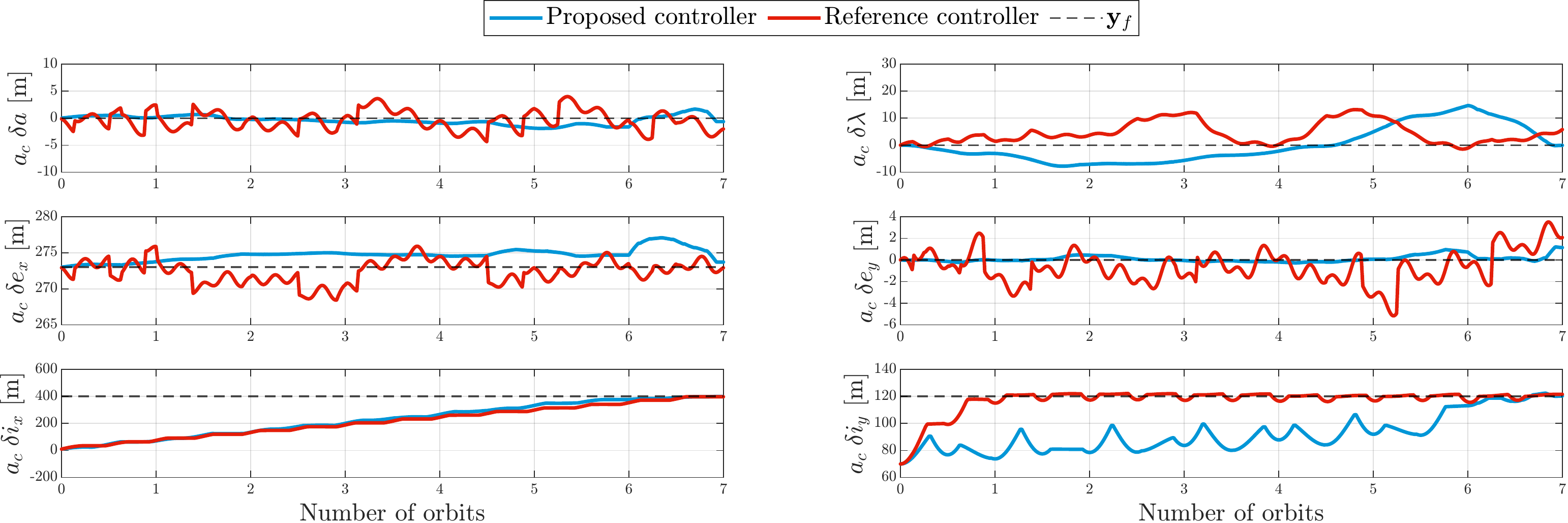}
    \caption{Closed-loop dimensional ROE profile}
    \label{fig:MPC_ROE_profile}
\end{figure}

The dimensional ROE profile generated by the proposed closed loop system as well as that which is generated by the reference control scheme (the one proposed in \cite{Belloni2023Relative}) are depicted in \cref{fig:MPC_ROE_profile}. It is clear that the two control algorithms could achieve the final required dimensional ROE vector, $\vec{y}_{f}$, at the end of the maneuver at approximately the same time.

\begin{figure}[ht]
    \centering
    \includegraphics[width=\linewidth]{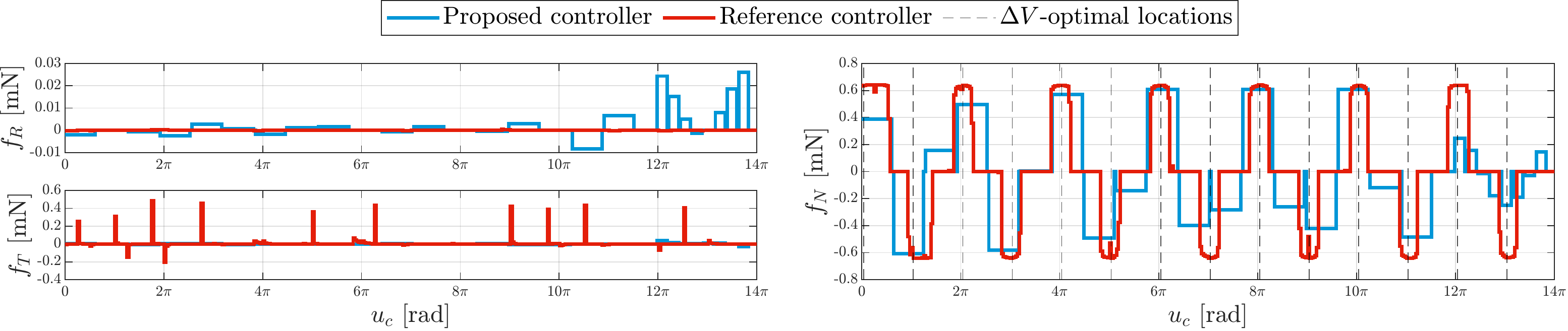}
    \caption{Closed-loop thrust profile in the RTN frame}
    \label{fig:MPC_thrust_profile}
\end{figure}

A comparison of the thrust profiles, projected to the RTN frame, between the proposed and the reference controllers is presented in \cref{fig:MPC_thrust_profile}. Since the maneuver is an OOP one, only thrust in the normal direction is expected to be exerted, however, one can notice that the reference control algorithm does occasionally fire in the transversal direction. The proposed as well as the reference controllers can be seen to provide almost no thrust in the radial direction. While this happens in the reference controller because a hard constraint is imposed on the radial thrust to be exactly zero, the proposed approach does not use radial thrust simply because it is optimal to not use it, as expected for this type of simulation scenarios. Furthermore, in \cref{fig:MPC_thrust_profile} the two controllers are shown to distribute the available thrust around the $\Delta V$-optimal locations since they are both $\Delta V$-optimal algorithms.\\
Two key performance metrics are compared for both controllers, the proposed and the reference, in \cref{tab:Benchmark_comparison}. The tables suggests that although the proposed algorithm uses more total $\Delta V$, $\Delta V_\text{tot}$, than the reference one, the terminal error of the proposed controller is much less than that of the reference.    
{\renewcommand{\arraystretch}{\arraystretchfortable}
\begin{figure}[ht]
    \begin{minipage}{0.55\textwidth}
    \centering
    \captionof{table}{Comparison between the proposed and the reference MPCs}
    \begin{tabular}{lcc}
        \hline
        \hline
        ~ & Terminal $\vec{y}$ error $[\text{m}]$ & $\Delta V_\text{tot} \;[\text{m}/\text{s}]$ \\
        \hline
        Proposed & {\setlength{\arraycolsep}{\hseparationofmatrixintable}\renewcommand{\arraystretch}{\arraystretchformatrixintable} $\begin{bmatrix}
        -0.6 &  -0.1 &  0.7 &  1.1 &  -0.3 &  0.3 
        \end{bmatrix}^{\intercal}$} & $0.65$\\
        Reference & {\setlength{\arraycolsep}{\hseparationofmatrixintable} \renewcommand{\arraystretch}{\arraystretchformatrixintable} $\begin{bmatrix} 3.6 &  -9.2 &  1.4 &  -2.0 &  2.9 &  -1.6 \end{bmatrix}^{\intercal}$} & $0.5$\\
        \hline
        \hline
    \end{tabular}
    \label{tab:Benchmark_comparison}
    \end{minipage}
    \hfill    
    \begin{minipage}{0.4\textwidth}
    \centering
    \includegraphics[width=\linewidth]{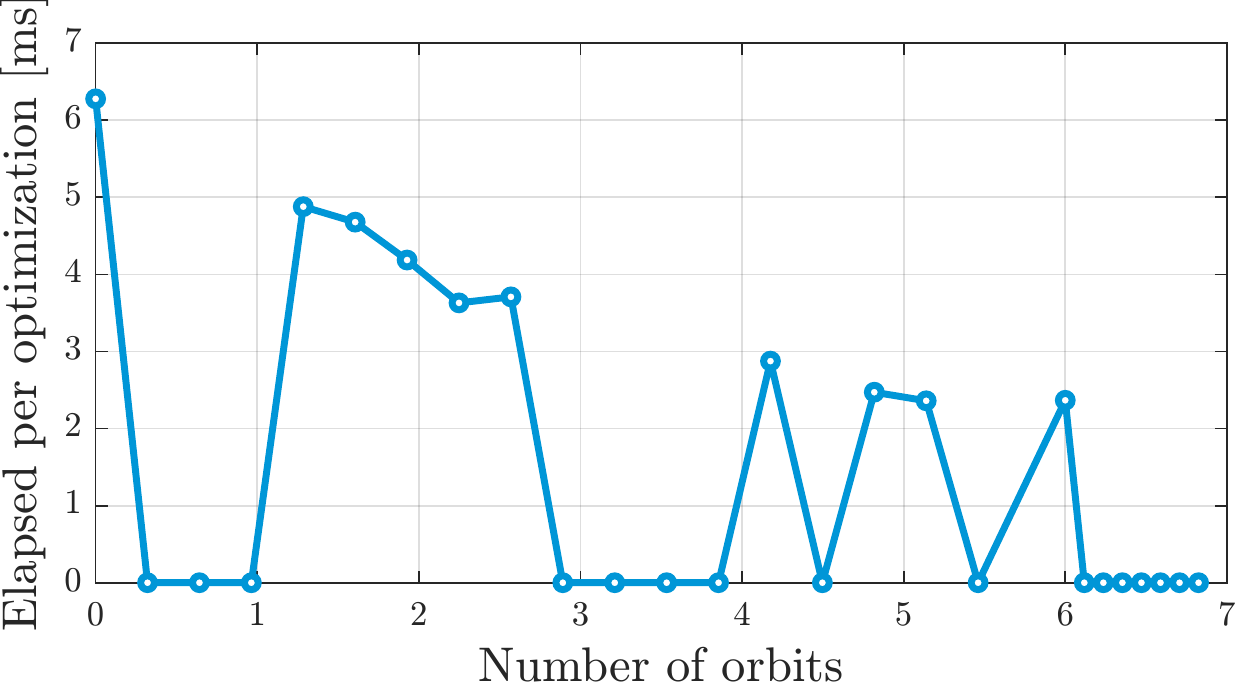}
    \captionof{figure}{Elapsed time to solve problem \ref{prob:QP_formulation}}
    \label{fig:MPC_elapsed}
    \end{minipage}
\end{figure}
}
Since the proposed control scheme is working much like an MPC, state prediction over the prediction horizon needs to be done recurrently through the guidance function, however, the guidance does not need to be run at each optimization step (see \cref{fig:Closed-loop-relative-simulation}). It is evident in Algorithm \ref{alg:Sim-closed-loop-logic} that the guidance is run only when the distance (in the dimensional ROE space) between current state and its prediction by the previous guidance profile is more than the threshold, $\epsilon$. To show how often the guidance problem needs to be solved over the benchmark simulation, the elapsed time for every guidance run is depicted in \cref{fig:MPC_elapsed}, where an elapsed time of exactly zero means that the guidance is skipped and the previous guidance profile is used. It is important to note that the elapsed time in \cref{fig:MPC_elapsed} correspond to the time it takes to construct and solve the problem using the OSQP solver \cite{osqp} interfaced with Matlab and run on an 8-core Intel Core i9-10885H processor. It is clear from \cref{fig:MPC_elapsed} that the elapsed time for every guidance run is getting smaller as the simulation advances since the size of the problem is getting smaller as the guidance problem is always solved from the current to the final time.

\section{Conclusion}
This note proposes a novel approach for guidance of an underactuated spacecraft to perform relative orbit corrections with respect to a reference satellite. The guidance scheme not only considers the typical constraints of the main spacecraft while performing the maneuver, but it also considers the dynamical constraint which arises from the satellite being equipped with a single electric thruster. Thanks to parameterizing the relative dynamics between the two spacecraft using the quasi non-singular relative orbital elements, the guidance problem was formulated as a quadratic programming problem, which, if feasible, is guaranteed to have exactly one solution that can be found using high-performance standard quadratic programming solvers.
The proposed guidance scheme shows numerous inherent merits which include fuel-optimality and the ability to support long no-thrust periods arising from operational constraints.
The paper also proposes a model-predictive-control-like scheme to close the control loop which does not require the guidance optimization to run at the beginning of each prediction horizon. This closed loop system has been validated via high fidelity numerical simulations in which navigation and actuators' errors and constraints are emulated, and its performance is benchmarcked against that of a reference controller from the literature. The proposed guidance and control algorithms have shown superior performance over the reference controller for the benchmark simulation in terms of terminal tracking errors.

\section*{Acknowledgments}
This research was funded in whole, or in part, by the Luxembourg National Research Fund (FNR), grant reference BRIDGES/19/MS/14302465. For the purpose of open access, and in fulfilment of the obligations arising from the grant agreement, the author has applied a Creative Commons Attribution 4.0 International (CC BY 4.0) license to any Author Accepted Manuscript version arising from this submission.\\

\bibliography{References.bib}

\end{document}